\documentclass[nofootinbib,twocolumn,showpacs,superscriptaddress,prd]{revtex4-1}
\usepackage{graphicx,amsmath,amssymb,dcolumn,bm}
\usepackage{epic,eepic}
\usepackage[dvipdfmx]{pict2e}
\usepackage{color}

\definecolor{mycolor}{rgb}{0.6,0.0,0.4}

\begin{document}
\preprint{KEK-TH-1954, J-PARC-TH-0083}
\title{Tensor-polarized structure function {\boldmath$b_1$} \\
by convolution picture for deuteron}
\author{W. Cosyn}
\affiliation{Department of Physics and Astronomy, Ghent University, 
             Proeftuinstraat 86, B9000 Ghent, Belgium}
\author{Yu-Bing Dong}
\affiliation{Institute of High Energy Physics, 
             Chinese Academy of Sciences, Beijing 100049, China}
\affiliation{Theoretical Physics Center for Science Facilities (TPCSF), 
             CAS, Beijing 100049, China}
\author{S. Kumano}
\affiliation{KEK Theory Center,
             Institute of Particle and Nuclear Studies, KEK \\
             1-1, Ooho, Tsukuba, Ibaraki, 305-0801, Japan}
\affiliation{J-PARC Branch, KEK Theory Center,
             Institute of Particle and Nuclear Studies, KEK, \\
           and
           Theory Group, Particle and Nuclear Physics Division, 
           J-PARC Center, \\
           203-1, Shirakata, Tokai, Ibaraki, 319-1106, Japan}
\author{M. Sargsian}
\affiliation{Department of Physics, Florida International University, 
             Miami, Florida 33199, USA} 
\date{February 9, 2017}
\begin{abstract}
There are polarized structure functions $b_{1-4}$ for the spin-1 deuteron. 
We calculated the leading-twist tensor structure function $b_1$ 
by using convolution description for the deuteron. 
We found large differences between our theoretical
functions and HERMES experimental data on $b_1$.
Although higher-twist effects should be considered in obtaining
experimental $b_1$, it suggests a possible existence of 
new hadron physics mechanism for spin-1 hadrons.
Furthermore, we found that there are significant distributions 
at large Bjorken $x$. In future, an experimental measurement is planned 
at JLab for $b_1$ and there is a possibility of a proton-deuteron Drell-Yan
experiment at Fermilab with the tensor-polarized deuteron, 
so that further theoretical studies are needed for clarifying 
the physics origin of tensor structure in terms of quark and 
gluon degrees of freedom.
\end{abstract}
\maketitle

\section{Introduction}
\label{intro}

The discovery of nucleon spin puzzle created the field of high-energy 
spin physics. So far, the studies have been focused on spin-1/2 nucleon.
It is known that there exist additional structure functions $b_{1-4}$
for a spin-1 hadron in charged-lepton deep inelastic scattering 
due to the existence of tensor structure 
\cite{fs83,hjm89}.
There are some theoretical studies on the tensor-polarized 
structure functions, and the first measurement of $b_1$ 
was reported by the HERMES collaboration \cite{Airapetian:2005cb}. 
A useful parametrization was proposed for the tensor-polarized
parton distribution functions (PDFs) from the HERMES data
\cite{Kumano:2010vz} by using a constraint for the $b_1$ sum 
proposed for the tensor-polarized valence-quark
distributions \cite{Close:1990zw}.
Although the $b_1$ data are not still accurate, the measurement
indicates interesting features different from conventional theoretical
calculations. As the deviation from the naive-quark-model prediction
created the topic of nucleon spin puzzle, it is important to show
a conventional theoretical estimate with the standard deuteron model
since a possible experimental deviation may indicate 
a ``tensor-polarization puzzle". 

As such a ``standard" model of the deuteron for describing
the twist-two structure function $b_1$, we use a convolution 
picture. Namely, the tensor-polarized structure function $b_1$ 
is given by unpolarized structure functions convoluted with
tensor-polarized lightcone momentum distributions of the nucleon.
There are such studies for $b_1$ in Refs. \cite{hjm89,Khan:1991qk}.
We try to test their results by independent ways.
We use two theoretical descriptions. One is a basic convolution model 
and another is the virtual-nucleon-approximation model.
Such convolution models have been used also as a baseline calculation 
for the nuclear structure function $F_{2}^A$ at medium and large $x$
in which nuclear effects were taken into account through the binding 
and nucleon Fermi motion in a nucleus together 
with short-range correlations.
These physics ingredients are contained in the spectral function, 
which is the nucleon's four-momentum distribution in the nucleus. 
We directly apply such descriptions for calculating the structure 
function $b_1$. If the calculated $b_1$ distribution is much different
from the HERMES measurement, a new hadron physics mechanism could
be considered beyond the standard model of the deuteron
in describing the tensor structure in terms of quark and
gluon degrees of freedom \cite{Miller:2013hla}.

The purpose of our research is to show the convolution result
on $b_1$ as the standard theoretical estimate and to discuss
its comparison with the HERMES data \cite{CDKS2017}. 
In particular, a new accurate measurement of $b_1$ will start 
in a few years at JLab \cite{Jlab-b1} and a possible Drell-Yan experiment 
is considered at Fermilab with a tensor-polarized deuteron
\cite{hk-1998,Kumano:2016ude}.
Furthermore, other experiments are possible in principle
for the tensor-polarized structure functions 
at BNL, EIC, J-PARC, GSI-FAIR, and IHEP@Russia.

\section{Theoretical formalisms for $\bm{b}_{\bm{1}}$}
\label{formalism}
\vspace{-0.3cm}


The cross section for deep inelastic scattering of charged lepton
from a spin-1 hadron is given by a hadron tensor multiplied
by a lepton tensor, and the hadron tensor is expressed 
in terms of eight structure functions as
\cite{hjm89,Kumano:2014pra}
\begin{align}
W_{\mu \nu}^{\lambda' \lambda} (P,q)
= &  \int \frac{d^4 \xi}{4 \pi} \, e^{i q \cdot \xi} \,
               \langle \, P, \lambda' \, | \, [ \, J_\mu^{\, em} (\xi) ,
                      J_\nu^{\, em} (0) ]  \, | \, P, \lambda \, \rangle
\nonumber \\[-0.15cm]
= & -F_1 \hat{g}_{\mu \nu} 
     +\frac{F_2}{M \nu} \hat{P}_\mu \hat{P}_\nu 
     + \frac{ig_1}{\nu}\epsilon_{\mu \nu \lambda \sigma} q^\lambda S^\sigma  
\nonumber \\[-0.15cm]
& +\frac{i g_2}{M \nu ^2}\epsilon_{\mu \nu \lambda \sigma} 
      q^\lambda (P \cdot q \, S^\sigma - S \cdot q \, P^\sigma )
\notag \\[-0.15cm]
&   -b_1 r_{\mu \nu} 
     + \frac{1}{6} b_2 (s_{\mu \nu} +t_{\mu \nu} +u_{\mu \nu}) 
\notag \\[-0.15cm]
&
     + \frac{1}{2} b_3 (s_{\mu \nu} -u_{\mu \nu}) 
     + \frac{1}{2} b_4 (s_{\mu \nu} -t_{\mu \nu}) .
\label{eqn:w-1}
\end{align}
The tensor-polarized structure functions are $b_{1-4}$,
which do not exist for the spin-1/2 nucleon. 
Here, $\hat{g}_{\mu \nu}$ and $\hat{P}_\mu$ 
are defined as
$ \hat{g}_{\mu \nu} \equiv  g_{\mu \nu} - q_\mu q_\nu / q^2 $ and
$ \hat{P}_\mu \equiv P_\mu - q_\mu P \cdot q / q^2 $
to satisfy the current conservation 
$q^\mu W _{\mu \nu} = q^\nu W _{\mu \nu}=0$.
The coefficients $r_{\mu \nu}$, $s_{\mu \nu}$, 
$t_{\mu \nu}$, and $u_{\mu \nu}$ are defined 
by the spin-1 polarization vector, 
hadron and virtual-photon momenta ($P$, $q$),
and initial and final spin states ($\lambda$, $\lambda'$),
and their actual expressions should be found in 
Refs.\,\cite{hjm89,Kumano:2014pra}.
The structure functions $b_3$ and $b_4$ are twist-4, and
$b_1$ and $b_2$ are leading-twist functions which are related
to each other by the Callan-Gross type relation $2xb_1 = b_2$
in the Bjorken scaling limit. We may first investigate the leading
function $b_1$ (or $b_2$).

The structure function $b_1$ is expressed in terms of 
the tensor-polarized PDFs $\delta_{_T} f (x,Q^2)$ 
in the parton model as
\begin{align}
b_1 (x,Q^2) & = \frac{1}{2} \sum_i e_i^2 
      \, \left [ \delta_{_T} q_i (x,Q^2) 
      + \delta_{_T} \bar q_i (x,Q^2)   \right ] , 
\nonumber \\
\! \!
\delta_{_T} f (x,Q^2) & \equiv f^0 (x,Q^2)
          - \frac{f^{+1} (x,Q^2) +f^{-1} (x,Q^2)}{2}.
\label{eqn:b1-parton}
\end{align}
Here, $f^\lambda$ is an unpolarized parton distribution
in the hadron spin state $\lambda$ and
$e_i$ is the charge of the quark flavor $i$.
The Bjorken scaling variable $x$ is defined as
$x =Q^2/(2 M_N \nu)$ with the nucleon mass $M_N$
and $\nu = P\cdot q /M$, and its range is given by
$0<x<2$ for the deuteron.

There is a sum rule for $b_1$ \cite{Close:1990zw,Kumano:2010vz}
in the similar way with the Gottfried sum rule \cite{flavor3}:
\begin{align}
\! \! \!
\int dx \, b_1 (x) 
    & = - \lim_{t \to 0} \frac{5}{24} \, t \, F_Q (t) 
\nonumber \\
  & \ \ \ 
     + \frac{1}{9} \int dx
      \, \left [ \, 4 \, \delta_{_T} \bar u (x) +  \delta_{_T} \bar d (x) 
                     +   \delta_{_T} \bar s (x)  \, \right ] ,
\nonumber \\
 \int \frac{dx}{x}
 \, [F_2^p & (x) - F_2^n (x) ] 
   =  \frac{1}{3} 
   +\frac{2}{3} \int dx \, [ \bar u(x) - \bar d(x) ] ,
\label{eqn:b1-sum-gottfried}
\end{align}
where $F_Q(t)$ is the electric quadrupole form factor
for the spin-1 hadron and the first term, which 
comes from tensor-polarized valence-quark distributions,
vanishes: $\lim_{t \to 0} \frac{5}{24} t F_Q (t)=0$.
It could be used as a guideline in investigating $b_1$.
As the Gottfried-sum-rule violation initiated the studies of $\bar u - \bar d$
\cite{flavor3}, a finite $b_1$ sum could indicate tensor-polarized
antiquark distributions. In fact, a finite $b_1$ sum was suggested
in the HERMES experiment, and it is interesting to measure 
the tensor-polarized antiquark distributions in 
the Fermilab-E1039 Drell-Yan experiment \cite{Kumano:2016ude}.

For calculating $b_1$ in the standard deuteron model with D-state
admixture, we introduce two convolution models. 
(A) One is a basic convolution description and 
(B) another is a virtual nucleon approximation 
    which includes higher-twist contributions.

\vspace{-0.3cm}
\subsection{Theory 1: Basic convolution description}
\label{convolution}
\vspace{-0.3cm}

In the convolution description of nuclear structure functions $W_{\mu\nu}^A$,
the hadron tensor is given by the nucleonic one $W_{\mu\nu}$ convoluted
with the momentum distribution of the nucleon, so called spectral function
$S (p)$, as
\vspace{-0.20cm}
\begin{align}
& \! \! \! \!
W_{\mu\nu}^A (P_A, q)  = \int d^4 p \, S(p) \, W_{\mu\nu} (p, q) ,
\nonumber \\
& \! \! \! \! 
S (p) = \frac{1}{A} \sum_i
      | \phi_i (\vec p \,) |^2 
     \delta \left ( p^0-M_A + \sqrt{M_{A-i}^{\ 2} 
                    + \vec p^{\ 2}} \, \right ) ,
\label{eqn:w-convolution}
\vspace{-0.20cm}
\end{align}
where $p$ and $P_A$ are momenta for the nucleon and nucleus,
and $\phi_i (\vec p \,)$ is the momentum-space wave function 
for the $i$-th nucleon. It explains major features of nucleon modifications
at medium and large $x$ ($x>0.2$) by the mechanisms of nuclear binding, 
Fermi motion, and short-range correlations contained 
in the spectral function.

The hadron tensor for the deuteron can be expressed by their helicity 
amplitudes of the virtual photon as \cite{hjm89}
\begin{align}
A_{hH,hH} (x,Q^2)  
   =  \varepsilon_h^{*\mu} \varepsilon_h^{\nu} \, 
       W_{\mu\nu}^D (p_{_D},q),
\label{eqn:A-helicity}
\end{align}
and the corresponding one $\hat A_{hs,hs} (x,Q^2)$ for the nucleon.
Here, $\varepsilon_h^{\mu}$ is the photon polarization vector.
The structure function $b_1$ of the deuteron and $F_1$ 
of the nucleon ($F_1^N$) are expressed as \cite{hjm89,kk08}
\begin{align}
b_1 & = A_{+0,+0} - \frac{A_{++,++} +A_{+-,+-}}{2} ,
\nonumber \\
F_1^N & = \frac{ A_{+\uparrow,+\uparrow} +A_{+\downarrow,+\downarrow} }{2} ,
\label{eqn:A-b1F1}
\end{align}
in the Bjorken scaling limit.
Using these equations, we obtain a convolution expression for $b_1$ 
defined by the one per nucleon as
\vspace{-0.20cm}
\begin{align}
b_1 (x,Q^2) & = \int \frac{dy}{y} 
\, \delta_T f(y) \, F_1^N (x/y,Q^2), 
\nonumber \\
\delta_T f(y) & \equiv f^0 (y)  - \frac{f^+ (y) + f^- (y)}{2} .
\label{eqn:b1-convolution}
\vspace{-0.20cm}
\end{align}
The lightcone momentum distribution is expressed 
by the deuteron wave function $\phi^H (\vec p \,)$ as
\vspace{-0.20cm}
\begin{align}
f^H (y) = \int d^3 p \, y \, | \, \phi^H (\vec p \,) \, |^2
          \, \delta \left ( y - \frac{E-p_z}{M_N}   \right ) .
\label{eqn:deuteron-momentum}
\vspace{-0.20cm}
\end{align}
The variable $y$ is the momentum fraction defined by
$ y   =  M \, p \cdot q / (M_{N} \, P \cdot q ) \simeq 2 \, p^- / P^- $
where the light-cone momentum is defined by
$\, p^- \equiv (p^0 -p^3)/\sqrt{2} \,$ by taking $z$-axis 
along the virtual-photon momentum direction.
Using the deuteron wave function with the D-state admixture, we obtain
\begin{align}
\delta_T f(y)  = \int d^3 p \, y &
 \left [ - \frac{3}{4 \sqrt{2} \pi} \phi_0 (p) \phi_2 (p) 
  + \frac{3}{16\pi} |\phi_2 (p)|^2 \right ]
 \nonumber \\
 & \times
 (3 \cos^2 \theta -1) \, \delta \left ( y - \frac{p\cdot q}{M_N \nu}  \right ) ,
\label{eqn:delta-t-f}
\end{align}
where $\phi_0 (p)$ and $\phi_2 (p)$ are S- and D-state wave functions.
According to this basic convolution model, the structure function $b_1$
arises due to the D-state admixture. 
For calculating Eq.\,(\ref{eqn:b1-convolution}), we need the structure 
function $F_1^N$. In our work, we use the leading-order (LO) expression
with the longitudinal-transverse ratio $R$ as
\begin{align}
F_1^N (x,Q^2) & = 
\frac{1+4 \, M_N^2 \, x^2/Q^2}{2 \, x \, [ 1+R(x,Q^2) ]}
     \, F_2^N (x,Q^2) ,
\nonumber \\
F_2^N (x,Q^2)_{\text{LO}} & = x \sum_i e_i^2 
     \left [ q_i (x,Q^2) + \bar q_i (x,Q^2) \right ]_{\text{LO}} .
\label{eqn:f1-lo}
\end{align}

\vspace{-0.5cm}
\subsection{Theory 2: Virtual nucleon approximation}
\label{vna-model}
\vspace{-0.3cm}

The cross section for the charged-lepton DIS from the polarized deuteron
is expressed by the polarization factors and structure functions:
\vspace{-0.00cm}
\begin{align}
\frac{d\sigma}{dx \, dQ^2} 
= & \frac{\pi y^2\alpha^2}{Q^4(1-\epsilon)}
\bigg[ F_{UU,T}+\epsilon F_{UU,L}
\nonumber  \\ 
& \ \ \ \ \ \ 
+T_{\parallel\parallel} \left( F_{UT_{LL},T}+\epsilon F_{UT_{LL},L} \right) 
\nonumber  \\ 
& \ \ \ \ \ \ 
+ T_{\parallel\perp} \cos\phi_{T_\parallel} \sqrt{2\epsilon(1+\epsilon)} \,
   F_{UT_{LT}}^{\cos\phi_{T_\parallel}} 
\nonumber  \\ 
& \ \ \ \ \ \ 
+ T_{\perp\perp} \cos(2\phi_{T_\perp}) \, \epsilon \,
   F_{UT_{TT}}^{\cos(2\phi_{T_\perp})}\bigg ]\,,
\label{eq:cross}
\vspace{-0.15cm}
\end{align}
where $\epsilon$ is the degree of the longitudinal polarization
of the virtual photon, the details on the longitudinal and transverse 
polarization factors
($T_{\parallel\parallel}$, $T_{\parallel\perp}$, $T_{\perp\perp}$)
and the angles ($\phi_{T_\parallel}$, $\phi_{T_\perp}$)
should be found in Ref.\,\cite{CSW}.
Among the structure functions, $ F_{UT_{LL},T} $
and $ F_{UT_{TT}}^{\cos(2\phi_{T_\perp})} $ are related to $b_1$
and they are expressed by the helicity amplitudes and tensor-polarized
structure functions $b_{1-3}$ as
\vspace{-0.15cm}
\begin{align}
F_{UT_{LL},T} & 
     =\frac{2}{\sqrt{6}} \left(A_{++,++}-2A_{+0,+0}+A_{+-,+-}\right) ,
\nonumber \\[-0.10cm]
& = 
-\frac{1}{x}\sqrt{\frac{2}{3}}\left[2(1+\gamma^2)xb_1-\gamma^2\left(\frac{1}{6}
b_2-\frac{1}{2}b_3\right)\right ] ,
\nonumber \\
F_{UT_{TT}}^{\cos(2\phi_{T_\perp})} & =-\sqrt{\frac{2}{3}}
\Re e A_{+-,-+} 
  = -\sqrt{\frac{2}{3}}\frac{\gamma^2}{x} 
\left(\frac{1}{6}b_2-\frac{1}{2}b_3 \right) ,
\label{eqn:futll-lt}
\vspace{-0.00cm}
\end{align}
where $\gamma =\sqrt{Q^2}/\nu$.
From these equations, $b_1$ is expressed by these two structure functions as
\begin{equation} 
b_1=- \frac{1}{1+\gamma^2} \sqrt{\frac{3}{8}} \,
   \left [ F_{UT_{LL},T}
    +F_{U\mathcal{T}_{TT}}^ { \cos(2\phi_ { T_\perp }) } \right ] .
\label{eq:b1_f}
\vspace{-0.00cm}
\end{equation}

Now, we explain the virtual nucleon approximation (VNA) for calculating $b_1$.
It considers the $np$ component of the light-front deuteron wave function.
The virtual photon interacts with one nucleon which is off the mass shell,
while the second non-interacting ``spectator'' is assumed to be
on its mass shell. Then, the structure functions are obtained by 
integrating over all possible spectator momenta $\vec p_N$:
\vspace{-0.15cm}
\begin{equation}
W_{\mu\nu}^{\lambda' \lambda} (P,q)
 \! = \! 4(2\pi)^3 \! \! \int \! d\Gamma_N
\frac{\alpha_{_N}}{\alpha_i} 
W_{\mu\nu} (p_{i},q)
\rho_D(\lambda',\lambda) ,
\label{eq:htensorfinal}
\vspace{-0.15cm}
\end{equation}
where $d\Gamma_N$ is the phase space for the spectator nucleon.
The factor $4 (2\pi)^3$ comes from the definition of deuteron lightcone 
wave function, and the factor $\alpha_{_N}/\alpha_i$ exists because 
the hadron tensor $W_{\mu\nu}$ is for the nucleon with momentum $p_i$
instead of the nucleon at rest \cite{CSW}.
The lightcone momentum fractions are defined for 
the interacting ($i$) and spectator ($N$) nucleons as
$ \alpha_i= 2 \, p_i^- / P^-$ and
$ \alpha_{_N}= 2 \, p_N^- / P^-=2-\alpha_i$.
The deuteron density $\rho_D(\lambda',\lambda)$ is defined by
the deuteron wave function 
$\Psi^{D}_\lambda(\vec{k},\lambda'_N,\lambda_N)$ as
\vspace{-0.15cm}
\begin{equation}
\rho_D(\lambda',\lambda)
= \! \! \! \sum_{\lambda_{N},\,\lambda_{N}'} \! \! \!
\frac{[\Psi^{D}_{\lambda'}(\vec k,\lambda_{N}',\,\lambda_{N}) ]^\dagger
\Psi^{D}_\lambda(\vec k,\lambda_{N}',\,\lambda_{N})}{\alpha_{_N} \, \alpha_i} .
\label{eqn:rho-d}
\vspace{-0.2cm}
\end{equation}
The wave function $\Psi^{D}_\lambda$ is then expressed by
the S- and D-wave components $\phi_0$ and $\phi_2$.
Calculating the structure functions in Eq.\,(\ref{eqn:futll-lt})
and using the relation of Eq.\,(\ref{eq:b1_f}), we finally obtain
$b_1$ in the VNA model:
\vspace{-0.20cm}
\begin{align}
b_1(x, & Q^2) = \frac{3}{4(1+\gamma^2)} \int \frac{k^2}{\alpha_i} 
                 dk \, d(\cos\theta_k) 
\nonumber \\[-0.15cm]
& \times
\bigg[ F_{1}^N(x_i,Q^2) \left(6\cos^2\theta_k-2\right) 
\nonumber \\[-0.18cm]
& \ \ \ \ \ 
-\frac{T^2 } {2 \, p_i \cdot q } F_ { 2 }^N (x_i ,Q^2)
\left(5\cos^2\theta_k-1\right) \bigg]
\nonumber \\[-0.05cm]
& \times
\left[ - \frac{\phi_0 (k) \phi_2(k)}{\sqrt{2}}+\frac{\phi_2( k)^2}{4}\right] ,
\label{eq:b1_vna}
\end{align}
where $\theta_k$ is the angle between $\vec k$ and 
$\vec q$, and $T^\mu$ is defined by
$ T^\mu  = p_N^\mu+ q^\mu p_N\cdot q / Q^2 - L^\mu p_N\cdot L / L^2$ with
$ L^\mu=P^\mu+ q^\mu P\cdot q / Q^2$.
As obvious from the above derivation, the $b_1$ of the VNA model
includes higher-twist contributions, whereas the first basic model
of Eq.\,(\ref{eqn:b1-convolution}) was obtained by using the relation
(\ref{eqn:A-b1F1}) in the scaling limit.

\vspace{-0.2cm}
\section{Results}
\label{results}
\vspace{-0.3cm}

We show results on $b_1$ by integrating the expressions
in Eqs.\,(\ref{eqn:b1-convolution}) and (\ref{eq:b1_vna}).
For this numerical evaluation, we choose the PDFs for calculating
$F_2^N$, the longitudinal-transverse ratio $R$, and 
the deuteron wave function.
They are taken as MSTW2008 (Martin-Stirling-Thorne-Watt, 2008) 
leading-order (LO) parametrization, the SLAC-R1998 parametrization,
and the CD-Bonn wave function, respectively.

\begin{figure}[t]
\begin{center}
   \includegraphics[width=6.0cm]{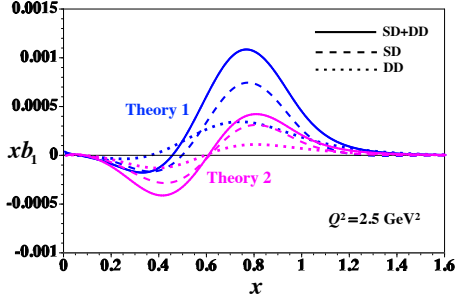}
\end{center}
\vspace{-0.6cm}
\caption{Structure function $b_1$ calculated 
by the two convolution descriptions at $Q^2$=2.5 GeV$^2$.
The contributions from the SD term, DD term, and their summation
are shown \cite{CDKS2017}.}
\label{fig:xb1-sd-dd}
\vspace{-0.0cm}
\end{figure}

Since the average scale of the HERMES measurement is $Q^2=2.5$ GeV$^2$,
we show our result at this $Q^2$. In Eqs.\,(\ref{eqn:b1-convolution}) 
and (\ref{eq:b1_vna}), there are two components, $\phi_0 \phi_2$
and $\phi_2 ^{\ 2}$, which are called SD and DD terms.
These contributions are shown in Fig.\,\ref{fig:xb1-sd-dd}
with total $b_1$ curves for the two theoretical descriptions.
The order of magnitude is rather small and the distributions 
are less than $10^{-3}$. We notice that these results are very
different from previous convolution estimates of
Refs.\,\cite{hjm89,Khan:1991qk} in the following points.
\vspace{-0.18cm}
\begin{itemize}
 \item[(1)] Although theoretical formalisms are similar, 
    our distributions, namely magnitude and $x$ dependence,
    are very different. Especially, the SD curves have opposite
    sign to the one in Ref.\,\cite{Khan:1991qk}.
\vspace{-0.18cm}
 \item[(2)] The finite distributions exist at large $x$, even 
    at $x>1$ although there is no distribution in Ref.\,\cite{Khan:1991qk}.
\end{itemize}
\vspace{-0.18cm}
There are also relatively large differences between two theory results. 
We checked that both results become similar in the scaling limit,
which indicates that higher-twist effects are the major sources
for the differences. In addition, there are effects coming
from slightly different normalizations for the lightcone wave functions
\cite{CDKS2017}.

\begin{figure}[t]
\begin{center}
   \includegraphics[width=6.0cm]{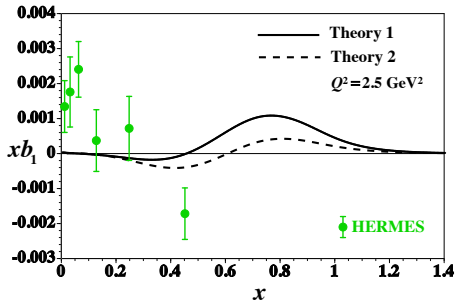}
\end{center}
\vspace{-0.6cm}
\caption{Calculated structure functions at $Q^2=2.5$ GeV$^2$
are compared with HERMES experimental data \cite{CDKS2017}.}
\label{fig:xb1-hermes}
\vspace{-0.3cm}
\end{figure}

Next, our results are compared with the HERMES data in 
Fig.\,\ref{fig:xb1-hermes}. It is obvious that theoretical
curves are much different from the data. In the measured
$x$ range ($x<0.5$), the experimental magnitude is one-order
larger than both theoretical estimates. 
Furthermore, there are relatively large distributions even 
at large $x$ ($0.6<x<0.8$). Because the HERMES errors are 
large, we cannot draw a solid conclusion from this comparison.
However, the large differences 
indicate that possibly a new hadron physics mechanism could 
be needed for their interpretation, although there are still some
rooms to improve, for example, by considering higher-twist
effects in extracting $b_1$ experimentally from 
the spin asymmetry $A_{zz}$ as pointed out in Ref.\,\cite{CDKS2017}.

It is puzzling to find the large differences between
the data and our standard convolution descriptions.
In future, the JLab experiment will start in a few years to
measure accurately $b_1$ at medium $x$ ($0.3<x<0.5$),
and there is a possibly to measure the proton-deuteron Drell-Yan
process in the Fermilab-E1039 Drell-Yan experiment \cite{Kumano:2016ude}.
Therefore, such a puzzle should be clarified by future studies;
however, further theoretical studies are needed to clarify the situation
and to consider a new mechanism to explain the HERMES data.
Possibly, a new hadron spin field could be explored by such studies.

\vspace{-0.4cm}
\section{Summary}\label{summary}
\vspace{-0.4cm}

We calculated the tensor-polarized structure function $b_1$ by using
the standard deuteron model with D-state admixture and the two convolution
models. We found that our $b_1$ values are much smaller in magnitude
than the HERMES data in the range $x<0.5$. It could indicate
possible existence of a new hadron mechanism for interpreting
the large differences, although other contributions such as higher-twists
should be investigated.

\vspace{-0.5cm}
\begin{acknowledgements}
\vspace{-0.4cm}

This work was supported by Japan Society for the Promotion of Science (JSPS)
Grants-in-Aid for Scientific Research (KAKENHI) Grant Number JP25105010.
\end{acknowledgements}



\end{document}